\documentclass[aps, twocolumn, a4paper, superscriptaddress, 10pt, pre, floatfix]{revtex4-2}
\usepackage{graphicx}
\usepackage{amsmath, amssymb}
\usepackage{lmodern}
\usepackage[T1]{fontenc}
\usepackage[utf8]{inputenc}
\usepackage[english]{babel}
\usepackage{xcolor}
\definecolor{oliver}{rgb}{0.0, 0.5, 0.0}

\begin{document}

\title{Quantifying scale-free behaviors in Rock-Paper-Scissors Models\\ as a function of Mobility}

\author{D. Bazeia}
\affiliation{Departamento de Física, Universidade Federal da Paraíba, 58051-970 João Pessoa, PB, Brazil}

\author{M. Bongestab}
\affiliation{Departamento de Física, Universidade Federal da Paraíba, 58051-970 João Pessoa, PB, Brazil}

\author{M. J. B. Ferreira}
\affiliation{Departamento de Física, Universidade Estadual de Maringá, 87020-900 Maringá, PR, Brazil}

\author{B. F. de Oliveira}
\affiliation{Departamento de Física, Universidade Estadual de Maringá, 87020-900 Maringá, PR, Brazil}

\author{J. E. B. Santos}
\affiliation{Departamento de Física, Universidade Estadual de Maringá, 87020-900 Maringá, PR, Brazil}
        
\begin{abstract}
    We investigate the scale-free behavior of the spatial rock-paper-scissors model with May-Leonard dynamics, analyzing specific quantifiers that engender the power-law feature. The main results show that an important parameter that drives the scale-free behavior is the mobility, which can be used to quantitatively describe several scale-free aspects of the model, such as the number of clusters, the characteristic length, the individuals' lifespan and its corresponding mean traveled distance. All of these are novel quantifiers of current practical interest for the study of biodiversity. 
\end{abstract}

\maketitle

\section{ Introduction } 

In nature, biodiversity plays a central role in the maintenance and stability of ecosystems \cite{leveque2004biodiversity, nowak2006evolutionary}.
In the past few decades, much effort has been dedicated in order to find the underlying mechanisms behind biodiversity. Unfortunately, the complexity of such systems make them hard to treat mathematically completely. However, many mathematical models have been proposed, in order to reproduce at least some specific aspects of systems that support biodiversity \cite{frean2001rock, Rev0, Rev1}. 

In the corresponding literature, a well established class of models refers to the so-called rock-paper-scissors (RPS) models. In its simplest form, the system is composed by three competing species interacting in a cyclic dominant nonhierarchical predator-prey scheme, where scissors dominates paper, paper dominates rock, and rock dominates scissors, mimicking the dynamics of the popular rock-paper-scissors game. In this symmetric model, no species has any advantage over the other. The main aspect of RPS models is the fact that cyclic dominance plays a crucial role in explaining biodiversity in nature \cite{sinervo1996rock,kerr2002local, kirkup2004antibiotic, reichenbach2007mobility, Liao}. In addition, it can be observed in many ecological systems as, for example: in blotched side lizards matting strategy \cite{sinervo1996rock}, amoeba \cite{shibasaki2018cyclic}, plant systems \cite{taylor1990complex, lankau2007mutual}, coral reefs \cite{jackson1975alleopathy}, salmon population oscillations \cite{guill2011three} and antibiotic mediated interaction between bacteria \cite{kirkup2004antibiotic}. See also Refs. \cite{Rev0,Rev1,lu2022effects,duan2024does,lu2024understanding,lu2025enhancement} for more information of current interest on models of evolutionary games based on rules engendering cyclic dominance.

As one knows, in a spatial RPS model the individuals are distributed over a region, which, for modeling purposes, is taken as a discrete square lattice with individuals taking place at its sites. In this work it will also be used periodic boundary conditions, which is a common practice in this type of simulations. Here we have considered the model governed by May-Leonard dynamics. This was first proposed analytically in \cite{may1975nonlinear} and numerically implemented, for instance, in \cite{reichenbach2007mobility}, where empty sites are allowed in the lattice. In this case, besides cyclic dominance, the individuals mobility are also crucial to the regulation of several features of the model, as we will see throughout this paper. In particular, biodiversity is jeopardized and may be lost when mobility exceeds a certain critical value \cite{reichenbach2007mobility}.

It is known from literature that besides supporting biodiversity, the May-Leonard dynamics also leads to the emergency of spatial structures that evolve engendering spiral shapes. In a qualitative analysis, it can be noticed that the diffusive process is fundamental to spiral patterns formation.
In particular, the spatial patterns as well as their size are closely related to the mobility parameter. Typically, the size of the spatial structures grows as the mobility parameter grows \cite{bazeia2022influence}, suggesting the mobility parameter plays the role of adjusting the scale of spiral sizes. For sufficient small values of mobility there will be no visible spiral formation. On the other hand, for high values of mobility, higher than the critical mobility, the system looses biodiversity. For intermediate values, however, the scale seems to depend only on the mobility parameter, irrespective to the lattice size. This is known from Ref. \cite{reichenbach2007mobility}, which also informs that RPS models exhibit a qualitative scale-free behavior with respect to the mobility parameter. In general, a system is said to exhibit scale-free behavior with respect to a parameter, $M$, if a given quantity $f(M)$ representing some property of the system obeys $f(\beta M) = g(\beta) f(M),$ being $\beta$ a scale parameter and $g(\beta)$ a continuous function. It can be shown that a function $f$ with such a property will always be a power-law;  see, \textit{e.g.}, Refs \cite{fisher1967theory, stanley1987introduction, barabasialbert}. In other words, a system is said to be scale-free with respect to $M$ if there is a quantity $f(M)$ such that
$f(M)= C M^{-\gamma},$
being $C$ a constant and $\gamma$ the power coefficient.

The present work brings quantitative analyses of some scale-free aspects of RPS model with May-Leonard dynamics. This is basically done by seeking for power-laws \cite{powerlaw} via distinct quantifiers, such as the number of clusters, characteristic length of the system, individuals lifespan and mean traveled distance, whose specific meanings will be better explained below. Each of these specific properties is assessed as a function of mobility, which can be used to specify regions in the parameter space such that the system behaves scale-free.

\section{Methodology}

Here we consider the spatial RPS model with May-Leonard dynamics, defined over a square lattice sized $N \times N$ with periodic boundary conditions. We consider the von Neumann lattice, in which each site has four neighbors, as we further explain below. To implement the numerical simulation, we suppose that a field $\phi_i$ is assigned to each site $i$ of the lattice, which represents either an empty space (denoted by $0$, with the color white) or an individual of species $1$, $2$ or $3$, denoted by blue, red and yellow, respectively. In this non-hierarchical cyclic dominant predation rule, species 1 dominates species 2, species 2 dominates species 3 and species 3 dominates species 1.

Spatially, each individual interacts only with its four nearest neighbors (up, down, left and right). In this dynamics the individuals interact by three different actions: mobility, reproduction and predation, where each action has its corresponding rate of choice $m$, $r$ and $p$, respectively, such that $m+r+p=1$. These rates are the free  parameters of the model. We have run several simulations with $r\neq p$ and we have noticed no important qualitative changes in the results obtained with $r=p$. For this reason we have adopted $r=p$ throughout the present study. This choice also make the results in closer comparison with previous works in literature, such as Ref. \cite{reichenbach2007mobility, bazeia2022influence}.

The mobility in RPS is a diffusive random walk process, where the average area explored by a single individual per unit of time is proportional to a mobility parameter $M$. This parameter is related to the mobility rate $m$, which is proportional to $2MN^{2}$ \cite{reichenbach2008self, zhang2009four, bazeia2017hamming, park2019fitness}, being $N$ the linear size of the lattice. It is important to clarify that the parameter $M$ is, in fact, independent of $N$. Using the normalization condition $m+r+p=1$ one can write all the rates in terms of the mobility parameter $M$ as
\begin{eqnarray}
    \label{eq-rates}
    m &=& \frac{MN^2}{1+MN^2},\\
    \label{eq-ratesrp} r = p &=& \frac{1/2}{1+MN^2}.
\end{eqnarray}
In other words, as done in \cite{reichenbach2007mobility, zhang2009four,park2019fitness}, $M$ is the rescaled mobility parameter of the simulation. Here, the mobility $M$ is a non-negative real number.

The time evolution of the dynamical system starts in a random initial configuration, generated by assigning a random individual ($0$, $1$, $2$ or $3$) to each site. At the end of such a process each occupation covers $1/4$ of the lattice on average. At each iteration, the following steps are taken:

\begin{enumerate}
    \item a site $i$ (called active) is randomly selected;
    \item a site $j$ (called passive) is randomly selected from the neighborhood of $i$;
    \item an action (mobility, reproduction or predation) is randomly selected.
\end{enumerate}

The next step is to implement the selected action, if possible. If mobility is chosen, then the individuals on both sites are exchanged, i.e.,  $(i,j) \rightarrow (j,i)$; this action is carried out whenever it is chosen. If reproduction is selected and the passive site is empty, then the reproduction occurs, creating a copy of the the active individual at the passive site $j$, i.e.,  $(i,0) \rightarrow (i,i)$. Otherwise, if the passive site is not empty, the active individual will be unsuccessful to reproduce in the passive site. Likewise, if predation is chosen, the cyclic predation rule must be satisfied in order for the active individual to predate the passive one; in this case $(i,j) \rightarrow (i,0)$ when $j$ is the prey of $i$. It is important to stress that the empty sites are never able to be active sites.

According to \cite{reichenbach2007mobility}, biodiversity is achieved in RPS model only for mobilities lower than a critical value $M_c=(4.5 \pm 0.5) \times 10^{-4}$. In this case the time evolution reaches a dynamical equilibrium with alternating dominance among the species after a certain relaxation time. For mobilities above this critical threshold, the system always evolves to a homogeneous state with two species going extinct while the remaining one occupies the whole lattice. Since our focus concerns studying properties in the biodiversity phase of the model, we examine the regime where species densities oscillate around a mean value, with the emergence of spiral patterns. This restricts the mobility to values lower than the critical value, more precisely $M \leq 2.0\times 10^{-4}$. Moreover, the relaxation time is the time necessary for the system to reach the dynamical equilibrium phase, which will occur only after a sufficient number of predation actions. Numerical tests reveal that the system consistently reaches the equilibrium phase after $10^8$ predation interactions with the abundance of the species oscillating around a well-defined mean value, with $M$ values below $2.0\times 10^{-4}$. The higher the mobility, the lower the predation rate (see Eq. \ref{eq-ratesrp}), meaning the system takes longer to reach the equilibrium phase as mobility increases. In other words, relaxation time depends on the mobility parameter $M$. For this reason, time will be measured in units of predation actions throughout this work.

This study investigates how macro and micro quantitative aspects of the RPS model vary in response to the mobility parameter $M$. Specifically, we examine how the number of clusters of same individuals, characteristic length, individuals lifespan and its mean traveled distance respond to changes of $M$. These two last quantities represents a novelty in the literature of spatial May-Leonard models. Each of the quantifiers used in this work is discussed in detail below.

Examining the formation of domains of individuals of the same species (clusters) in spatial models may be of particular interest. In this work we have considered that two individuals belong to the same cluster if both belong to the same species $X$ and there is a path in the lattice made only by individuals of species $X$ connecting them. It is expected that for small mobilities $(M \approx 10^{-7})$, individuals will have a short-range influence on others, leading to the formation of small clusters or, equivalently, the formation of a large number of small clusters. As $M$ increases, larger spatial structures start to emerge, making the size of the clusters bigger or, equivalently, making the number of clusters to decrease. In general, the number of clusters may depends on the mobility parameter in a non-linear way. We have used the percolation algorithm proposed in \cite{hoshen1976percolation} in order to count the number of clusters in our simulations. Preliminary simulations using random states have shows that the largest cluster contains, on average, $(45 \pm 5)$ individuals. For this reason we have only considered cluster with more than $50$ individuals in our analysis.

Similarly, another important quantity related to the size of spatial structure is the characteristic length, which is obtained by the auto-correlation function $C(s)$ that measures the correlation between individuals $s$ units of length apart. It has been used following the definition given in \cite{avelino2022lotka} for the auto-correlation function. In this case, let $\mathcal{L}$ represent the regular square lattice, and $\phi_i$ an integer field for each site $i$, where it can take values representing one agent of any species $1$, $2$, $3$ or an empty site $0$. The distance between two sites $i$ and $j$ in the taxi cab metric is represented by $d(i,j)$. For a given site $i$ and a number $s=0, 1, 2, \cdots , N$ we define a set $\Pi_s(i)$ by
\begin{equation}
    \Pi_s(i) = \{j \in \mathcal{L} \;/\; d(i,j)=s\Delta s\},
\end{equation}
where $\Delta s=1$ is the grid spacing of the regular square lattice. The auto-correlation coefficient between sites $s \; (\neq 0)$ units of distance apart is defined by
\begin{equation}
    f_s=\dfrac{1}{2}\dfrac{1}{(s+1)}\sum_{i\in \mathcal{L}}\phi_i\left( 
\sum_{j \in \Pi_s(i)} \phi_j \right),
\end{equation}
the factor $1/2$ is due to the fact that every pair of
sites are counted twice for $s \neq 0$. For $s = 0$, on the other hand, the auto-correlation coefficient is simply
\begin{equation}
    f_0=\sum_{i\in \mathcal{L}} \phi_i^2.
\end{equation}
The auto-correlation function is then defined by the quotient of these two coefficients
\begin{equation}
    C(s)= \dfrac{f_s}{f_0}.
\end{equation}
In particular, $C(0)=1$ meaning the auto-correlation of an individual with itself is unity. The auto-correlation function gives important information about the average length of influence one individual has on others. The typical characteristic length $\ell$ is defined as the distance required to obey $C(\ell) = 0.15$.

\begin{figure*}[!htbp]
    \centering
    \includegraphics[width=16.8cm]{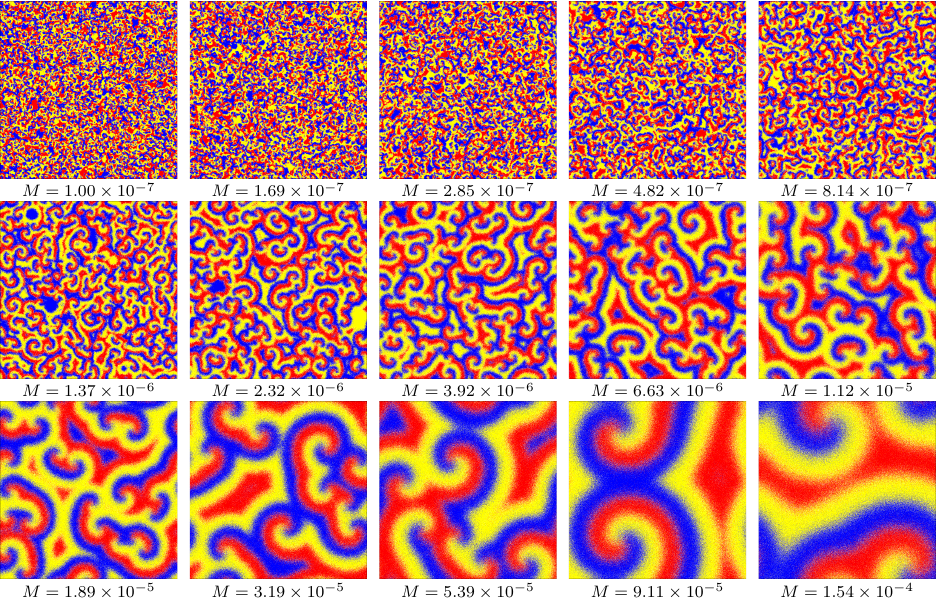}
    \caption{Snapshots of a lattice sized $N=10^3$ after $10^8$ predations as a function of the mobility parameter $M$. Here white dots represents empty spaces while the red, blue and yellow dots represent species $1, 2$ and $3$, respectively.} 
    \label{fig:1}
\end{figure*}

In addition, two others aspects of the RPS model are also visited in this work: individual lifespan and mean traveled distance. Here, by lifespan we mean the number of time steps an individual undergoes from birth until death on average, counted in number of predation actions the individuals participated in. The lifespan, as a global parameter, will also depends on the mobility $M$ in a non-linear way. It is also reasonable to think individuals will have a shorter life for small mobilities, since it increases the predation rate.

The mean traveled distance, on the other hand, is defined as the Euclidean distance between the point where an individual is born and the point where it dies, on average. This metric captures the spatial mobility of individuals throughout their lifetimes, and it is directly affected by mobility, offering insights into how far they disperse from their origin before vanishing. In this sense, the mean traveled distance reflects the impact of mobility on spatial dispersion while the lifespan serves as a quantifier of individual persistence. Both metrics provide valuable tools for analyzing the interplay between mobility, survival, and spatial movement in stochastic spatial simulations.

In this work, we have performed simulations of spatial RPS models monitoring the quantities discussed previously as a function of mobility. The results obtained along with the corresponding simulations parameters are presented below. The standard error of the results is of the order $10^{-5}$, lying within the points size and lines width used in the work. We refer to \cite{altman2005standard} for more details on standard error.

\section{Results}

We have set $N=10^3$ for all simulations performed. As said before, the usual relaxation time for such a lattice size is $10^8$ actions of predation actions. Thus all simulations were performed for over $2\times 10^8$ predation actions, being the first half (before the system reaches the dynamical equilibrium phase) discharged.

It is known from literature that biodiversity is jeopardized by mobility, the closer $M$ gets to the critical mobility  $M_c=(4.5 \pm 0.5) \times 10^{-4}$ \cite{reichenbach2007mobility}. For this reason, in Fig. \ref{fig:1} it is displayed the final state of the simulations for $M$ ranging from $1.0 \times 10^{-7}$ to $1.54 \times  10^{-4}$, below the critical value. It can be noticed from the panel the increase in size of the spatial structures as $M$ progressively increases indicating the loss of locality in the interactions. Also there is a visual clue that the stable phase of the systems gets more heterogeneous as $M$ increases, suggesting a more evident scale-free behavior of the system. Although this qualitative analysis can guides our investigation, in order to go further into the subject it is needed to  quantify the results in a proper way, as it is done below.

\begin{center}
{\it Number of Clusters} 
\end{center}

\begin{figure}[!htbp]
    \centering
    \includegraphics[width=8.4cm]{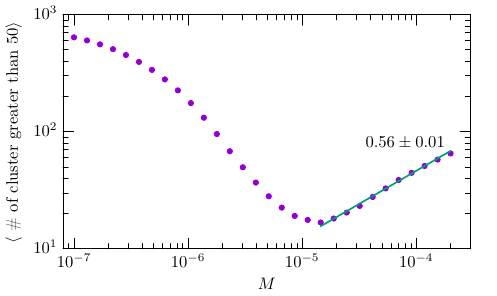}
        \caption{Number of clusters greater than $50$ individuals averaged over $10^3$ simulations as a function of $M$, for a lattice with $N=10^3$.} 
    \label{fig:2}
\end{figure}

On a square lattice of linear size $N=10^3$, the random initial configuration exhibits clusters with maximum size around $45 \pm 5$. For this reason, it has been only considered clusters greater than $50$ individuals. Thus in Fig. \ref{fig:2} it is depicted the number of cluster as a function of the mobility averaged over $10^3$ simulations. 

In the log-log scale it is possible to see the number of clusters decreasing non-linearly with $M$ until it reaches a minimum around $M = 1.5\times 10^{-5}$, and this is an expected result as discussed before. From this point on, however, the number of clusters grows as a straight line, indicating a power-law in the tail and a more heterogeneous system. In other words, the system behaves scale-free with respect to the number of clusters from this point on. The power-law coefficient, $\gamma_c$, in this case is estimated to be $\gamma_c = 0.56 \pm 0.01$. This value $M=1.5 \times 10^{-5}$ is the reference value which we used to depict the straight line in Fig. \ref{fig:2}; it will also be used for the other calculations in this work.

\begin{center}
{\it Characteristic Length} 
\end{center}

\begin{figure}[!htbp]
    \centering
    \includegraphics[width=8.4cm]{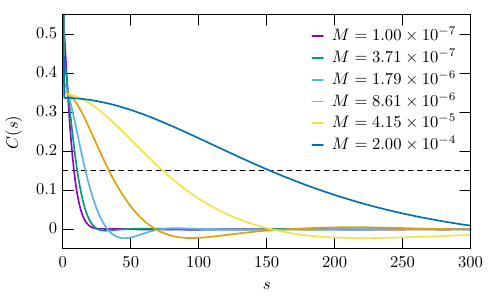}
    \caption{Auto-correlation function as a function of $M$ after $10^8$ actions of predation on a lattice sized $N=10^3$. This result was averaged over $10^3$ simulations.} 
    \label{fig:3}
\end{figure}

Let us now look at another important quantifier of the global aspect, the characteristic length. This quantity gives us important insights about the size of spatial structures. Fig. \ref{fig:3} shows the auto-correlation function as a function averaged over $10^3$ simulations, with the error of the line width used in this work. It can be seen in the picture that $C(r)$ has a general profile starting from $C(0)=1$, meaning total correlation between the focal individual and itself. The function then decays sharply at short distances, indicating the characteristic scale of local clustering, and asymptotically approaches zero as $r \rightarrow N$, meaning distant individuals are uncorrelated due to spatial heterogeneity.

\begin{figure}[!htbp]
    \centering
    \includegraphics[width=8.4cm]{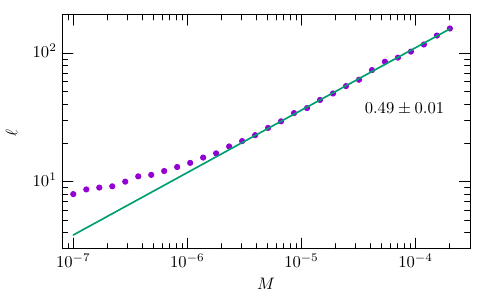}
    \caption{Characteristic length $\ell$ as a function of $M$ for a lattice with $N=10^3$ after $10^8$ actions of predation. This result was averaged over $10^3$ simulations.} 
    \label{fig:4}
\end{figure}

From $C(r)$, one can obtain the characteristic length such that $C(\ell)=0.15$. In Fig. \ref{fig:4}, the characteristic length is shown as a function of mobility. Again, there is no power-law behavior for mobility as low as $M=1.00\times 10^{-6}$. However, the more mobility increases, the more the spatial patterns become evident, and also the power-law appears in the tail.

This transition reflects the emergence of scale-free behavior, where the spatial correlation length displays a power-law dependence with mobility. Suggests that at higher mobility values, individuals become more distributed, while maintain correlations over large distances. For instance, when $M = 1.0 \times 10^{-4}$, an individual is correlated with other within a radius of $10^2$ sites. In comparison to the latticed used it is $10\%$ of its linear size.

It is possible to notice that the scale-free behavior observed via the characteristic length starts sooner in relation to the mobility parameter compared to the number of clusters. However, for comparison purposes, the fit displayed in Fig. \ref{fig:4} starts at the same point ($M=1.5\times 10^{-5}$). The estimated power-law coefficient in this case is $\gamma_l = 0.49 \pm 0.01$. This exponent value suggests that spatial structures grow sublinearly with increasing mobility, indicating a regime where mobility enhances spatial mixing but does not completely eliminate domain formation.

\begin{center}
{\it Individuals Lifespan}
\end{center}

The next quantifier to be considered is the individuals lifespan. This is a novel quantification procedure presented in the present work. It is a measurement at the individual level of the stochastic simulations. The quantification is done by keeping track of the individuals age from the moment of their birth (when they are reproduced) until their death (when they are predated), counted in terms of the actions of predation. This provides a direct measure of survival time under varying mobility conditions, allowing us to analyze how individual persistence is affected by $M$.

\begin{figure}[!htbp]
    \centering
    \includegraphics[width=8.4cm]{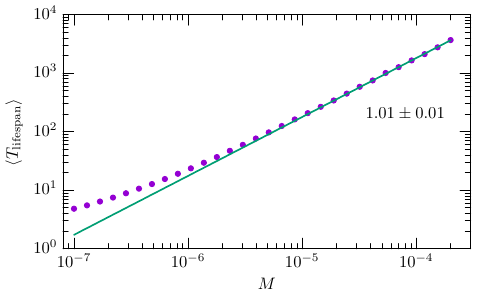}
    \caption{Individuals lifespan as a function of mobility in a lattice sized $N=10^3$. Here each point is averaged over $10^3$ simulations.}
    \label{fig:5}
\end{figure}

The numerical results are displayed in Fig. \ref{fig:5} as a function of mobility, averaged over $10^3$ simulations. From the picture it can be seen the power-law in the tail, with the estimated coefficient $\gamma_t = 1.01 \pm 0.01$. This scaling suggests that, beyond a critical mobility threshold, the probability distribution of lifespan follows a power-law profile. In the log-log plot shown on Fig. \ref{fig:5} the linear dependency with the mobility $M$ is clearly observed.

This result highlights that increased mobility not only alters spatial patterns but also affects the temporal persistence of individuals in the system. Furthermore, the observed power-law behavior in lifespan is consistent with other quantifiers in the model, such as characteristic length, reinforcing the presence of scale-free dynamics within the May-Leonard RPS model.

\begin{center}
{\it Mean Traveled Distance}
\end{center}

\begin{figure}[!htbp]
    \centering
    \includegraphics[width=8.4cm]{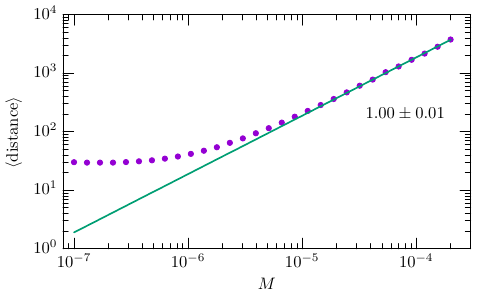}
    \caption{ Mean traveled distance as a function of mobility $M$ after $10^8$ predations in a lattice sized $N=10^3$. } 
    \label{fig:6}
\end{figure}

Simultaneously to the individuals' lifespan, the last quantifier considered is the mean traveled distance of a single individual during its lifetime. Here, one keeps track of the individual's steps taken from the position at birth until death, also counted in terms of the actions of predation. In Fig. \ref{fig:6}, one can see the mean traveled distance averaged over $10^3$ simulations as a function of mobility $M$.

As shown in the figure, it also exhibits a power-law in the tail, with the estimated coefficient $\gamma_d= 1.00 \pm 0.01$. This result indicates that the mean traveled distance scales linearly with mobility, meaning that, on average, individuals travel distances proportional to their mobility rate. The exponent value suggests a diffusive-like behavior, where displacement grows predictably with increasing mobility.

This power-law relationship strengthens even further the presence of scale-free dynamics in the system. In addition, the relationship between the traveled distance and the individuals' lifespan highlights how individuals' movement patterns are linked to survival time, reinforcing that mobility influences both spatial and temporal persistence in the May-Leonard RPS model.

\section{Ending comments}

In this work, we have studied macroscopic and microscopic aspects of the well known spatial RPS model under the May-Leonard stochastic rules. It was first shown that in a two dimensional lattice the number of clusters follow a power-law behavior for $M$ above a particular minimum value of the diffusive parameter, around $1.54 \times 10^{-5}$. This interesting result sheds light on self-similarity properties of the complex system under investigation. The computation of the correlation length confirms this observation and gives more information on the global aspect of the system, since it also displays a power-law behavior for $M$ above $10^{-5}$. We have also investigated the average individuals lifespan and the distance traveled through the time evolution. The main results are depicted in Figs. \ref{fig:2}, \ref{fig:4}, \ref{fig:5} and \ref{fig:6}.
We stress that the two new quantifiers, the lifespan and the distance traveled, are also of interest since they add information which may be of good use to describe the behavior of systems in the natural environment.

The study offers new perspectives for investigating global aspects and microscopic details of the nature of spatially structured interacting systems. While it is not possible yet to construct the dynamics from first principles, the broadening of this investigation may be of potential interest for real life situations. The study of cluster size of a competing community could shine light on the parameters of the dynamics of interest, and, on the other way around, life expectancy and average traveling distances can be inferred from numerical simulations, adding important information of current practical interest. These techniques can then be applied to many other systems, in particular, to the case of a larger number of species, as considered for instance in Refs. \cite{szabo2007segregation,peltomaki2008three,szabo2008self,wang2010effect,avelino2012junctions} and also, adding asymmetry in the rules that control the time evolution. They can be further studied in several other lattice based models, for instance, Public Good Games \cite{BO,hauert2002volunteering,chen2015competition,han2024selection}, fire spread \cite{perry1998current, sullivan2009wildland, schertzer2018fire}, Lotka-Volterra \cite{lotka1920, volterra1931theorie, Rev1} and also, in the case of off-lattice models \cite{JTheoBio,Arenzon2014,avelino2018spatial,bazeia2021SR}, to name some possibilities of current interest.

We notice that if one changes the May-Leonard rules considered in this work to the Lotka-Volterra ones, one needs to fuse the reproduction and predation rules into a single one, where the active site is now supposed to predate and reproduce simultaneously \cite{lotka1920,volterra1931theorie}. In this case there are no empty sites anymore, and the system is known to evolve differently, with a distinct clustering behavior and the absence of spirals, so the investigation of the above quantifiers may show distinct evolutions. Moreover, in the context of the Public Goods Games, the rules of evolution may be changed in several distinct ways \cite{BO}, so it is also of current interest to further examine the behavior of the above quantifiers under this new possibility of local interactions. Another possibility concerns RPS models that evolve under the action of asymmetric rules: in this case, an interesting result is known as the survival of the weakest \cite{SW}, which has also been identified in a microbial community of three distinct species (see \cite{Liao} and references therein) where an asymmetry may lead to unbalanced community that is dominated by the weakest species. In this case, the dynamics is somewhat different, so the study of the scale-free behavior is also of current interest.

We can also deal with other distinct lattice arrangements, including the kagome lattice, where two regular hexagons alternate with two equilateral triangles around each vertex, and the hexagonal lattice, which contains three neighbor, and the Kitaev hexagonal lattice, where the three neighbors now act differently; see, e.g., Refs. \cite{JRSI,RMP} for further details on these possibilities. We can also think of changing the topology of the system: in the square lattice with periodic boundary conditions, all the species interact living in a torus, but we can consider other possibilities, for instance, changing the torus to a spherical surface, and this could be studied to see if and how the modification could change the scale-free behavior found in the present investigation. We are investigating some distinct models, hoping to further report on these issues in the near future.

\begin{acknowledgments}
This work is supported by Conselho Nacional de Desenvolvimento Científico e Tecnológico (CNPq, Grants 303469/2019-6 (DB), 402830/2023-7 (DB), 309835/2022-4 (BFO)), Coordenação de Aperfeiçoamento de Pessoal de Nível Superior (CAPES, Grant 88887.688488/2022-00 (MB)) and Fundação de Apoio a Pesquisa do Estado da Paraíba (FAPESQ-PB, Grant 0015/2019).
\end{acknowledgments}

\end{document}